# Rare-Earth-Tuned Evolution from *d*- to *f*-Orbital Dominance and Giant Anomalous Hall Effect in Topological *R*GaGe (*R* = Ce, Pr, Nd) Semimetals

Zhian Xu[1†], Jian Yuan[1†,*], Ze Yan[1†], Xia Wang[1, 3], Na Yu[1, 3], Shihao Zhang[2*], and Yanfeng Guo[1,4*]

[1]State Key Laboratory of Quantum Functional Materials, School of Physical Science and Technology, ShanghaiTech University, Shanghai 201210, China

[2]School of Physics and Electronics, Hunan University, Changsha 410082, China

[3]Analytical Instrumentation Center, School of Physical Science and Technology, ShanghaiTech University, Shanghai 201210, China

[4]ShanghaiTech Laboratory for Topological Physics, ShanghaiTech University, Shanghai 201210, China

The family of noncentrosymmetric rare-earth germanides *R*GaGe (*R* = Ce, Pr, Nd) provides a rich materials platform to explore the intertwined physics of strong magnetism, electronic correlations, and topological band structures. Through a combination of crystal growth, characterization, and first-principles calculations, we reveal that these compounds exhibit a pronounced uniaxial magnetic anisotropy, leading to distinct ground states: *R*GaGe orders ferromagnetically with moments along the crystallographic *c*-axis, and shows an antiferromagnetic-like structure in the *ab*-plane. A key finding is a significantly enhanced intrinsic anomalous Hall conductivity (AHC) compared to their well-known *R*AlGe counterparts, which even reaches as high as 948 $\Omega^{-1}\cdot cm^{-1}$ at 2 K in PrGaGe. Our theoretical analysis predicts that this AHC originates from a robust Weyl semimetallic state driven by inversion symmetry breaking, where Weyl points near the Fermi level couple strongly to the magnetic order. Importantly, this topological state persists above the magnetic ordering temperature, confirming its intrinsic electronic origin. Our calculation also reveals that, while the near-Fermi-level states in CeGaGe and PrGaGe are dominated by *d*-orbital contributions, NdGaGe exhibits significant *f*-orbital involvement, signaling a progressive evolution from *d*- to *f*-orbital dominated topology. These results establish the *R*GaGe system as a tunable platform for systematically extending the *R*AlGe-related family, showcasing a large anomalous Hall response and orbital evolution near the Fermi level, and advancing the understanding of the interplay between topology and magnetism in quantum materials.

†Zhian Xu, Jian Yuan and Ze Yan contributed equally to this work.

*Corresponding author. Email: yuanjian@alumni.shanghaitech.edu.cn; zhangshh@hnu.edu.cn; guoyf@shanghaitech.edu.cn.

In recent years, research on materials with topological band structures has made remarkable advances. Efforts to unravel how symmetry governs the Berry curvature of electronic wavefunctions have fueled a continuous expansion of experimental investigations into novel quantum phenomena in topological materials [1–20]. It is well-established that breaking time-reversal symmetry (TRS) in magnetic materials can generate an intrinsic anomalous Hall effect (AHE) driven by Berry curvature. Similarly, in noncentrosymmetric materials, the breaking of spatial inversion symmetry is crucial for enabling a spectrum of emergent quantum phenomena, including nonlocal gyrotropic effects [21], quantum nonlinear Hall effect [22-24], nonlinear optical responses [25,26], and chiral anomaly transport [27-29].

This context establishes the unique significance of the ternary tetragonal compounds *R*Al*X* (*R* = rare earth, *X* = Si or Ge). Their noncentrosymmetric crystal structure (space group $I4_1md$) and the associated Weyl semimetal (WSM) state give rise to distinctive topological properties, capturing considerable attention [30–42]. Notably, LaAlGe has been identified via angle-resolved photoemission spectroscopy (ARPES) as a type-II Weyl semimetal [39]. Introducing magnetic rare-earth elements at the *R* site further enriches their physical behavior, leading to emergent magnetism and anomalous transport. For instance, CeAlSi is a noncollinear ferromagnetic (FM) WSM exhibiting an anisotropic AHE [33], while PrAlSi, a centrosymmetric ferromagnet, shows a magnetic-field-induced AHE along the crystallographic *c*-axis [31]. Coupled with pronounced magnetic anisotropy and strong spin-orbit coupling (SOC), these characteristics establish the *R*Al*X* family as a fertile platform for investigating the interplay between electronic topology and magnetic order, offering a promising avenue to discover novel correlated topological phases. Despite this progress, a systematic exploration of how strategic chemical substitution within this structural family modulates this interplay, and consequently its magnetotransport signatures, remains less developed. Addressing this gap is crucial for advancing the fundamental understanding and functional control of topological quantum phenomena.

In this work, we bridge this gap by systematically synthesizing single crystals of the *R*GaGe series (*R* = Ce, Pr, Nd), achieved by substituting Ga for Al in the *R*Al*X* framework, and conducting a thorough investigation of their transport and magnetic properties. We uncover a pronounced evolution of magnetic order: *R*GaGe displays ferromagnetism when the magnetic field is along the *c*-axis (*H* // *c*) and antiferromagnetic-like behavior when the field is within the *ab*-plane (*H* // *ab*). Strikingly, the *R*GaGe family possesses a substantially enhanced intrinsic anomalous Hall conductivity (AHC) compared to its *R*AlGe counterparts. First-principles calculations reveal that these materials host a robust WSM state due to inversion symmetry breaking, with Weyl points near the Fermi level contributing to a large AHC that persists above the magnetic ordering temperature. The calculated band structure and measured transport properties collectively support the existence of a topologically nontrivial electronic state. Furthermore, the electronic structure exhibits a strong rare-earth dependence: while CeGaGe and PrGaGe are dominated by *d*-orbital contributions near the Fermi level, NdGaGe shows significant *f*-orbital involvement indicative of enhanced spin-orbit splitting across the series. Our findings not only establish *R*GaGe as an excellent platform for observing topology-driven AHE but also reveal a tunable evolution from localized *d*- to *f*-orbital dominated topology, providing a systematic extension of the *R*AlGe-related family and highlighting the role of orbital character in the anomalous Hall response.

Single crystals of *R*GaGe were grown via the self-flux method using a mixed Ga-In flux. High-purity starting materials, including rare-earth metal pieces (abraded from 99.99% purity ingots), gallium particles (99.99%), germanium pieces (99.99%), and indium particles (99.99%), were weighed in a molar ratio of *R*:Ga:Ge:In = 1:2:1:10. The elements were loaded into an alumina crucible, which was then sealed under high vacuum (~ $10^{-5}$ Torr) inside a fused silica tube to prevent oxidation during high-temperature processing. The sealed ampoule was placed in a programmable box furnace. The synthesis procedure involved heating the mixture to 1100 °C at a controlled rate, maintaining this temperature for 24 hours to ensure complete homogenization and reaction, followed by slow cooling to 500 °C at a rate of 2 °C/h. At 500 °C, the residual liquid flux was separated from the grown crystals by centrifugation. Finally, the obtained crystals were extracted from the crucible. The resulting single crystals display a well-defined plate-like morphology with flat, shiny surfaces that are stable in air, as shown in the inset of Fig. 1(b).

Single-crystal X-ray diffraction (SXRD) was performed at room temperature using a Bruker D8 Venture diffractometer equipped with a Mo-$K_{\alpha 1}$ radiation source ($\lambda$ = 0.71073 Å). Magnetic properties, including DC susceptibility and isothermal magnetization, were measured using a commercial Magnetic Property Measurement System (MPMS-3, Quantum Design) over a temperature range of 2–300 K under various applied magnetic fields. Electrical and magnetotransport measurements, such as resistivity and Hall effect, were carried out using a commercial 9 T DynaCool Physical Property Measurement System (PPMS, Quantum Design).

First-principles calculations were performed within the framework of density functional theory using the generalized gradient approximation, as implemented in the Vienna *ab initio* simulation package (VASP) together with the projector augmented-wave method [43–47]. In all calculations, a Hubbard parameter $U$ = 5 eV was applied to the *f*-orbitals of the rare-earth atoms (*R* = Ce, Pr, Nd). The *d*- and *f*-orbitals of the *R* atoms as well as the *p*-orbitals of Ga and Ge atoms were used to project Bloch states onto Wannier functions. Subsequently, the WannierTools package was employed to compute the intrinsic anomalous Hall conductivity using a 101×101×101 *k*-point mesh.

The crystal structure of *R*GaGe is depicted in Fig. 1(a). Phase identification and assessment of crystal quality were performed via room-temperature SXRD. As shown in Fig. 1(c), the diffraction pattern obtained for PrGaGe was successfully indexed to a tetragonal system (space group $I4_1md$, No. 109) with lattice parameters $a = b$ = 4.260 Å and $c$ = 14.521 Å, and angles $\alpha = \beta = \gamma$ = 90°. These values align well with previously reported data for PrGaGe, confirming the structural consistency of the synthesized crystal [48]. Furthermore, the corresponding reciprocal space map reveals a well-defined lattice with no detectable impurity peaks or diffuse scattering, indicating high phase purity and excellent crystalline quality of the investigated sample.

Fig. 1(b) presents the temperature-dependent in-plane electrical resistivity, $\rho(T)$, measured for *R*GaGe (*R* = Ce, Pr, Nd) single crystals between 2 K and 300 K. All compounds exhibit metallic behavior down to 2 K. Distinct anomalies are observed near the characteristic temperatures $T_C$ = 6.74 K, 17.9 K, and 9.72 K for $R$ = Ce, Pr, and Nd, respectively. These resistivity drops correspond to the onset of magnetic ordering or a possible electronic transition, consistent with the magnetic susceptibility measurements presented in Figs. 2(a)–(c).

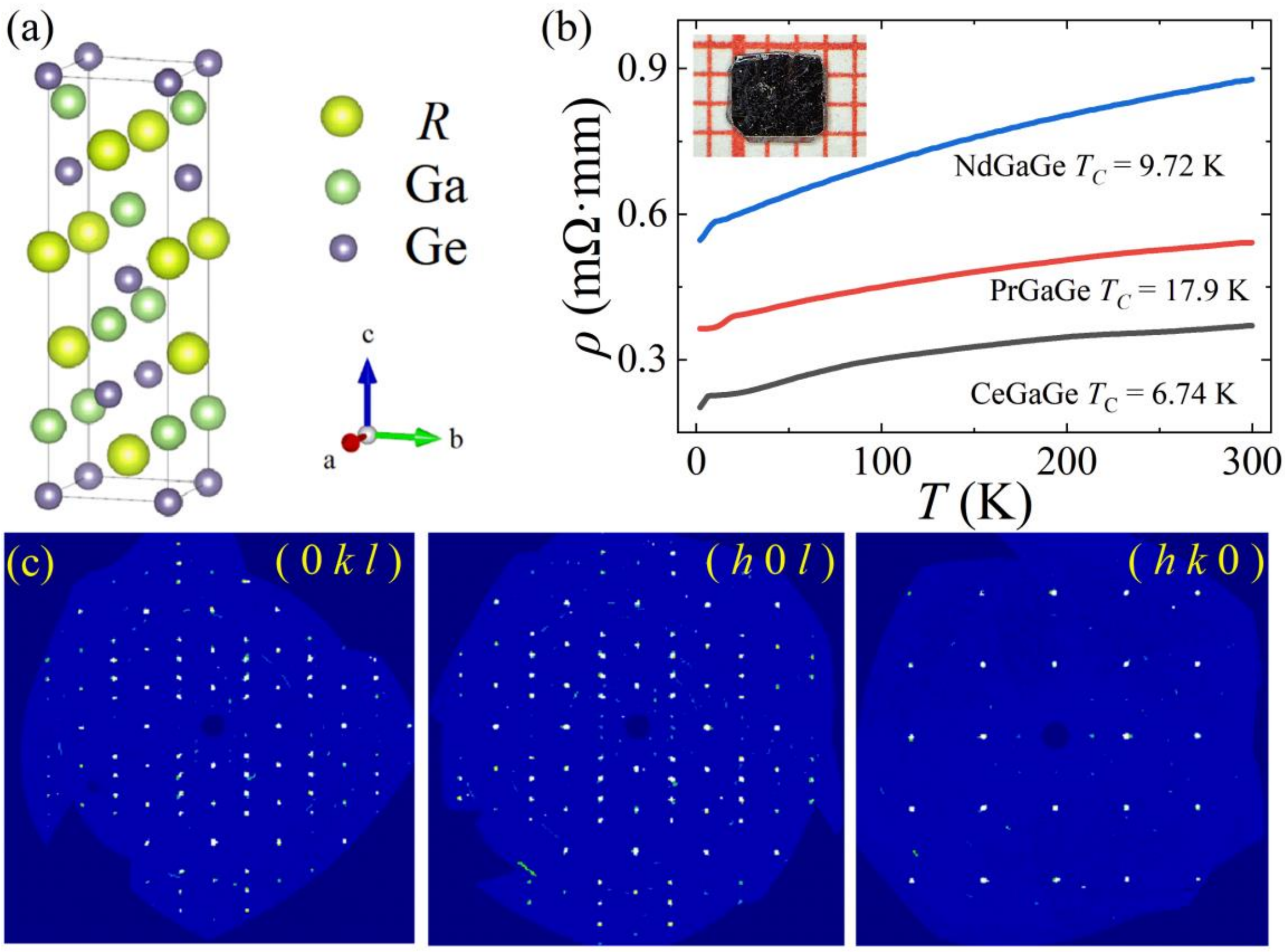

**Fig. 1.** (a) Crystal structures of $R$GaGe ($R$ = Ce, Pr and Nd). (b) Temperature dependence of the longitudinal resistivity $\rho$(T). The inset shows the image of a typical single crystal of PrGaGe. (c) Diffraction patterns of PrGaGe in the reciprocal space along (0$kl$), ($h$0$l$), and ($hk$0) directions.

To investigate the magnetic properties and spin anisotropy of the $R$GaGe single crystals, we performed comprehensive magnetization measurements as functions of both temperature and magnetic field. Temperature-dependent magnetic susceptibility was measured under zero-field-cooling (ZFC) and field-cooling (FC) conditions at $\mu_0H$ = 1 kOe, with the applied field aligned parallel to the crystallographic $c$-axis ($\chi_c$, $\mu_0H$ // $c$) and within the $ab$-plane ($\chi_{ab}$, $\mu_0H$ // $ab$), as shown in Figs. 2(a)–(c). In the $\chi$–$T$ curves, $\chi_c$ exhibits a clear paramagnetic-to-ferromagnetic (PM–FM) transition characterized by a sharp upturn as the temperature approaches the Curie temperature $T_C$, while $\chi_{ab}$ shows a distinct peak at $T_C$. The $\chi_c$ value of $R$GaGe saturates at $T_C$, and a peak appears in $\chi_{ab}$. We fit the reciprocal magnetic susceptibility 1/$\chi$($T$) for $H$ // $c$ using the Curie–Weiss law $\chi$(T) = $C/(T - \theta_P)$ in the temperature range of 50–300 K, where $C$ and $\theta_P$ denote the Curie constant and Weiss temperature, respectively. The fit yields $\theta_P$ values of approximately 5.19, 19.58, and 7.48 K for $R$ = Ce, Pr, and Nd, as shown in Figs. 2(a)–2(c), indicating FM interaction along the $c$-axis.

Additionally, the magnetization along the $c$-axis for $R$GaGe increases rapidly, with critical fields of 0.2, 0.3, and 0.35 T for $R$ = Ce, Pr, and Nd, respectively, suggesting FM behavior along the $c$-axis. Moreover, the $M(H)$ relationship shows a step-like transition at 2 T, which is quite similar to the antiferromagnetic (AFM) transition phenomenon in the $ab$-plane. The magnetic structure of $R$GaGe may involve non-collinear or tilted spin arrangements. Further neutron diffraction experiments are needed to determine the exact magnetic ground state. These pronounced drops observed in the susceptibility curves correspond closely to the anomalies in the electrical transport data, highlighting a strong coupling between magnetic order and electronic transport in these compounds.

The saturation moments obtained from high-field magnetization isotherms at 2 K in the polarized FM state align with the theoretical effective magnetic moments of the corresponding trivalent rare-earth ions: ~2.54 $\mu_B$/f.u. for $Ce^{3+}$, ~3.58 $\mu_B$/f.u. for $Pr^{3+}$, and ~3.62 $\mu_B$/f.u. for $Nd^{3+}$. This close agreement confirms that the localized moments are well described by Hund's rules and indicates a stable 3+ valence state for the rare-earth ions in the $R$GaGe structure.

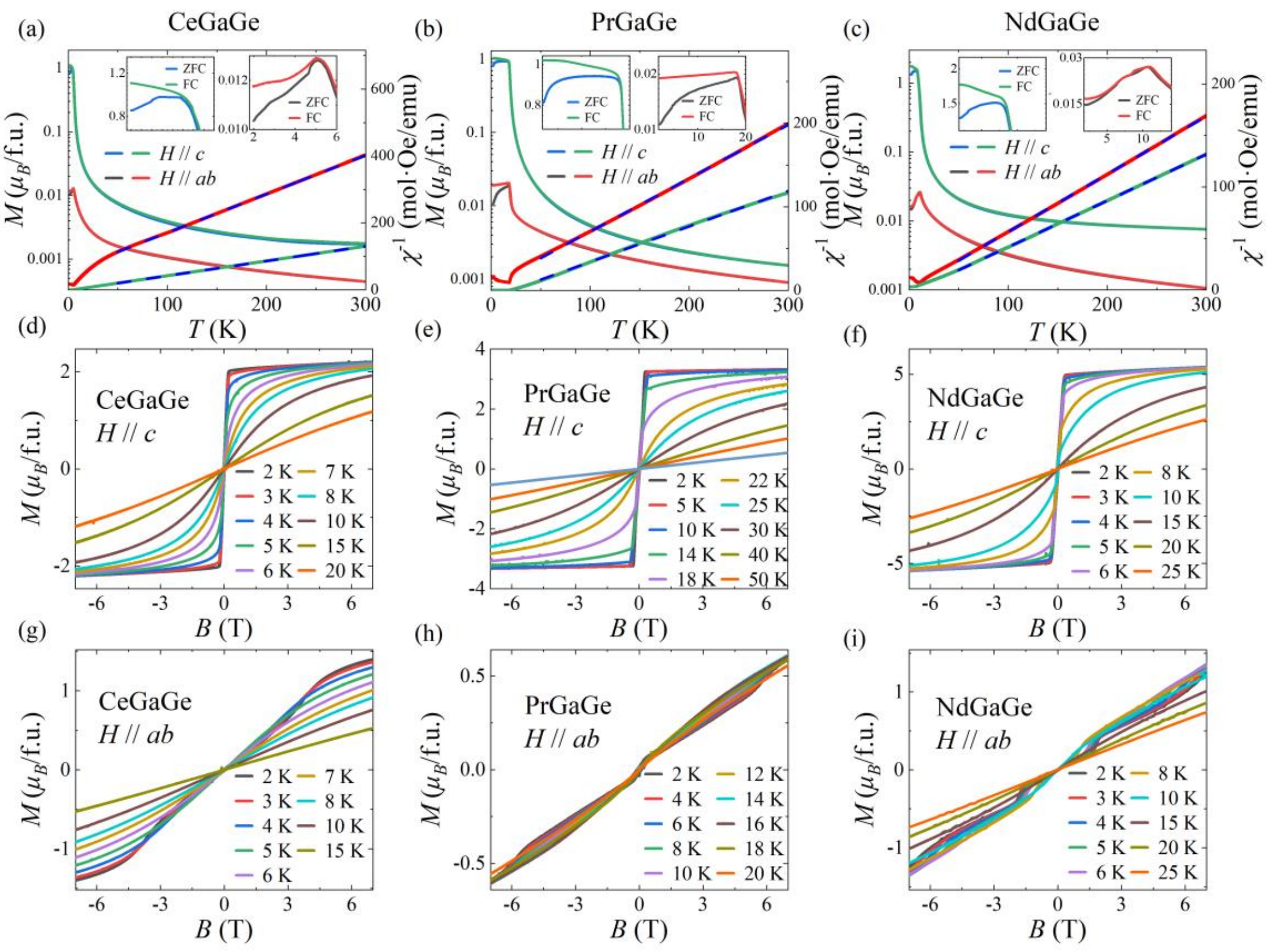

**Fig. 2.** (a–c) Left: Temperature-dependent magnetic susceptibility. Right: Curie-Weiss fit to the inverse susceptibility ($1/\chi$) of $R$GaGe measured at 0.1 T with $H$ // $c$ and $H$ // $ab$. The insets show an enlarged view of the low-temperature susceptibility behavior. (d–i) Magnetization as a function of applied magnetic field measured at various temperatures: panels (d–f) correspond to the field applied parallel to the $c$-axis, while panels (g–i) correspond to the field applied within the $ab$-plane.

The $R$GaGe ($R$ = Ce, Pr, Nd) series exhibits exceptionally strong uniaxial magnetocrystalline anisotropy. This is quantitatively reflected in the magnetization ratio $\chi_c/\chi_{ab}$, which is approximately 10 in the paramagnetic regime above $T_C$ and increases dramatically by nearly two orders of magnitude to ~ 100 at the base temperature of 2 K. The pronounced anisotropy originates from the synergistic interplay between the strong intrinsic SOC of the rare-earth ions and the low-symmetry crystalline electric field (CEF) imposed by the tetragonal structure (space group $I4_1md$). The unfilled $4f$ shells of the trivalent rare-earth ions retain unquenched orbital angular momentum, which, via strong SOC, gives rise to a large total angular momentum $J$—forming the fundamental basis for the directional dependence of the magnetic moment.

The tetragonal CEF lifts the degeneracy of the $(2J+1)$-fold ground multiplet, splitting it into a set of Stark levels and selecting a highly anisotropic ground-state wavefunction. This CEF-defined ground state strongly favors magnetization alignment along the crystallographic $c$-axis, i.e., the easy axis. The single-ion anisotropy induced by the CEF is already evident in the paramagnetic state, as seen in the susceptibility ratio $\chi_c/\chi_{ab}$ ~ 10 above $T_C$. Upon cooling through the magnetic ordering transition, the establishment of long-range ferromagnetic order collectively aligns these pre-existing anisotropic single-ion moments. This collective alignment dramatically amplifies the macroscopic anisotropy, leading to the greatly enhanced ratio $\chi_c/\chi_{ab}$ ~ 100 at 2 K. This evolution highlights how the single-ion CEF effect dictates the primary anisotropy axis, which is subsequently reinforced and stabilized by collective exchange interactions in the ordered state. Therefore, the exceptionally large and temperature-dependent magnetocrystalline anisotropy in *R*GaGe directly manifests the cooperative effect between local CEF physics and emergent magnetic order.

The pronounced uniaxial anisotropy originates from the cooperative effect of the single-ion CEF anisotropy, which defines the easy axis, and the exchange interactions that stabilize long-range magnetic order. Although CEF alone favors moment alignment along the $c$-axis, the actual magnetic ground state also depends on the sign and strength of exchange couplings, which may vary across the rare-earth series. This is consistent with observations in isostructural compounds such as CeAlSi (easy-plane FM) [33] and CeAlGe (AFM) [36], where exchange interactions modify the CEF-driven anisotropy.

Given the strong interplay between magnetism and charge transport in these materials, we systematically performed magnetotransport measurements, as presented in Fig. 3. The magnetoresistance (MR), measured with the magnetic field applied parallel to the $c$-axis and defined as $\mathrm{MR} = [\rho_{xx}(B) - \rho_{xx}(0)]/\rho_{xx}(0) \times 100\%$, where $\rho_{xx}$ denotes the longitudinal resistivity, shows a pronounced evolution across the ferromagnetic–paramagnetic phase boundary that correlates strongly with the magnetic ground state of CeGaGe, PrGaGe, and NdGaGe.

As shown in Figs. 3(a)–(c), the MR behavior varies markedly with temperature. Above $T_C$, a significant negative MR is observed; it peaks near the magnetic transition and weakens at higher temperatures. This feature is attributed to the suppression of critical spin fluctuations by the external magnetic field—a hallmark of strong magnetic correlations in the paramagnetic regime. Below $T_C$, the MR exhibits a nonmonotonic field dependence. A sharp negative MR emerges at low fields, arising from domain alignment and the suppression of magnon scattering [49]. At high fields, once the magnetization is nearly saturated, the MR turns positive due to the dominant orbital effect of the Lorentz force.

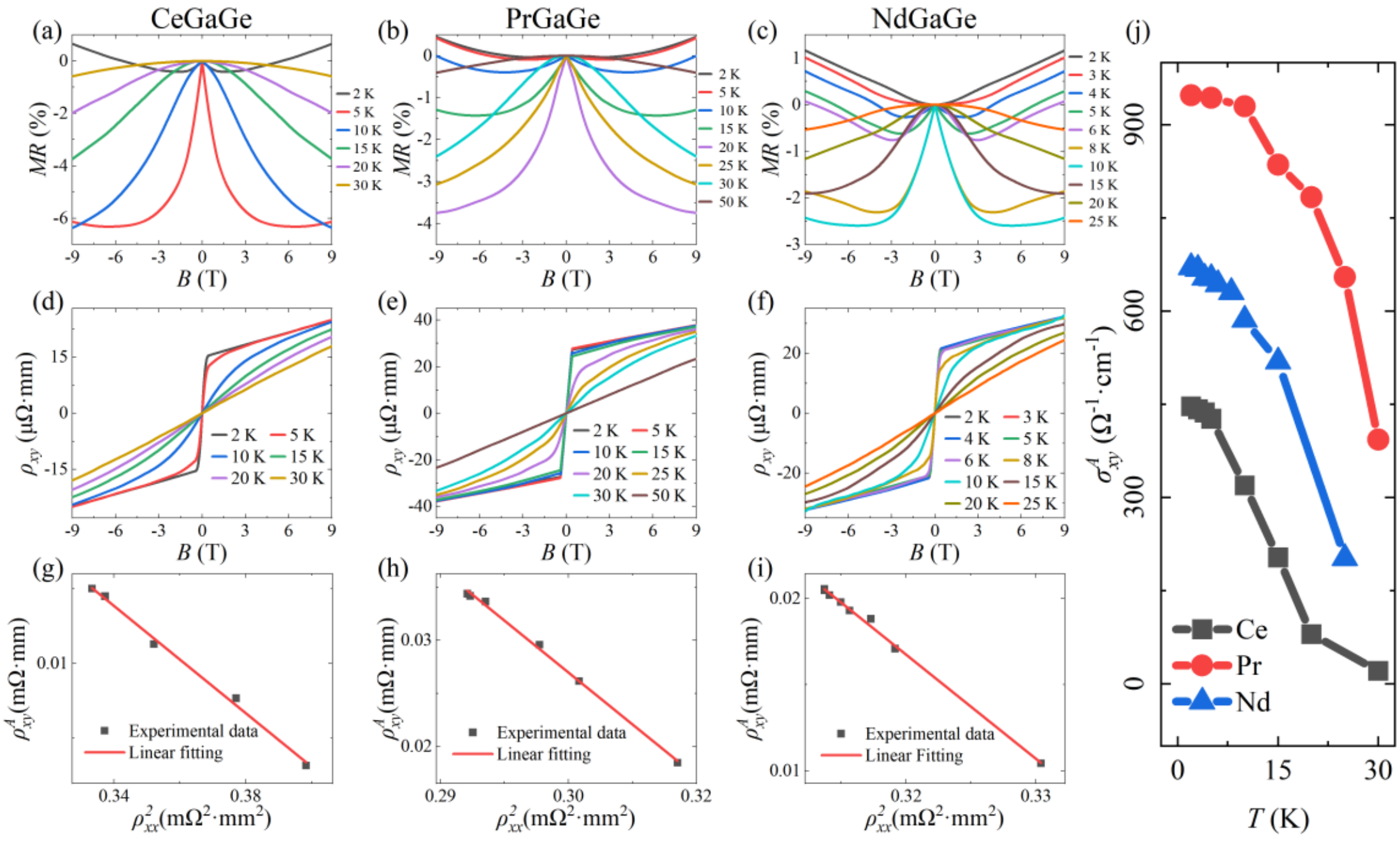

**Fig. 3.** (a–c) The temperature dependent MR and (d–f) Hall resistivity versus magnetic field $B$ at different temperatures for $R$ = Ce, Pr, and Nd, respectively. (g-i) Plots of $\rho^A_{xy}$ against $\rho_{xx}$, which represent the fitting curve by a linear function. (j) The calculated results of intrinsic anomalous Hall conductivities $\sigma^A_{xy}$. MR and (d–f) Hall resistivity ($\rho_{xy}$) as functions of magnetic field at various temperatures for $R$ = Ce, Pr, and Nd, respectively. (g–i) Plots of the $\rho^A_{xy}$ versus $\rho_{xx}$, fitted with a linear function. (j) The corresponding calculated $\sigma^A_{xy}$.

The anomalous Hall resistivity $\rho_{xy}$ of $R$GaGe is shown in Figs. 3(d)–(f). The anomalous Hall component $\rho_{AH}$ was extracted from the total Hall resistivity using the relation $\rho_{\mathrm{H}} = \rho^N{}_H + \rho^A{}_H = R_0B + 4\pi R_sM$ at various temperatures, where $\rho^N{}_H$ denotes the normal Hall resistivity. To determine its underlying origin, we employed the Tian–Ye–Jin model [50,51], which expresses as $\rho^A{}_H = \alpha\rho_{xx}(0) + a\rho_{xx} + \beta\rho^2_{xx}(0) + b\rho^2_{xx}$. Here, the coefficient $a$ denotes extrinsic skew-scattering contribution and the coefficient $b$ encompasses contributions from both intrinsic Berry curvature and side-jump mechanism. As shown in Figs. 4(g)–(i), anomalous Hall resistivities are linearly dependent on $\rho^2_{xx}$, which excludes extrinsic skew-scattering contribution. Skew scattering scales linearly with $\rho_{xx}$ and typically dominates in highly conductive systems with low residual resistivity, whereas the residual resistivities in $R$GaGe (10 ~ 50 μΩ·cm) fall within the moderately dirty regime where intrinsic and side-jump contributions are more relevant. The side-jump contribution can be estimated as $\sigma_{side\text{-}jump} \sim (e^2/hc)(\varepsilon_{\mathrm{SO}}/E_{\mathrm{F}})$, with $\varepsilon_{\mathrm{SO}}/E_{\mathrm{F}} \sim 0.01$ in FM metals [52], yielding an estimated side-jump conductivity ~ 2.7 Ω⁻¹·cm⁻¹ for $R$GaGe, which is two orders of magnitude smaller than the measured AHC. Given that the intrinsic contribution substantially exceeds the side-jump contribution across a wide range of scattering strengths, the dominant contribution must arise from the intrinsic Berry curvature. The resulting intrinsic AHC values are summarized in Fig. 3(j). At 2 K, the maximum AHC reaches 446, 948, and 670 Ω⁻¹·cm⁻¹ for $R$ = Ce, Pr, and Nd, respectively. Notably, the AHC of PrGaGe reaches 948 $\Omega^{-1}\cdot cm^{-1}$—approximately 1.3 times larger than the previously reported value for PrAlGe (~ 738 $\Omega^{-1}\cdot cm^{-1}$) [53]. These large AHC values indicate the presence of topologically nontrivial fermions in *R*GaGe. It is worth noting that the AHC decreases gradually upon warming but remains finite well above $T_C$. For example, in PrGaGe, $\sigma_{xy}^{A}$ at 30 K (~393 $\Omega^{-1}\cdot cm^{-1}$) is still ~ 41% of that value at 2 K. This confirms that the Berry curvature responsible for the AHE is not entirely quenched by the loss of long-range magnetic order, consistent with an intrinsic band-structure origin.

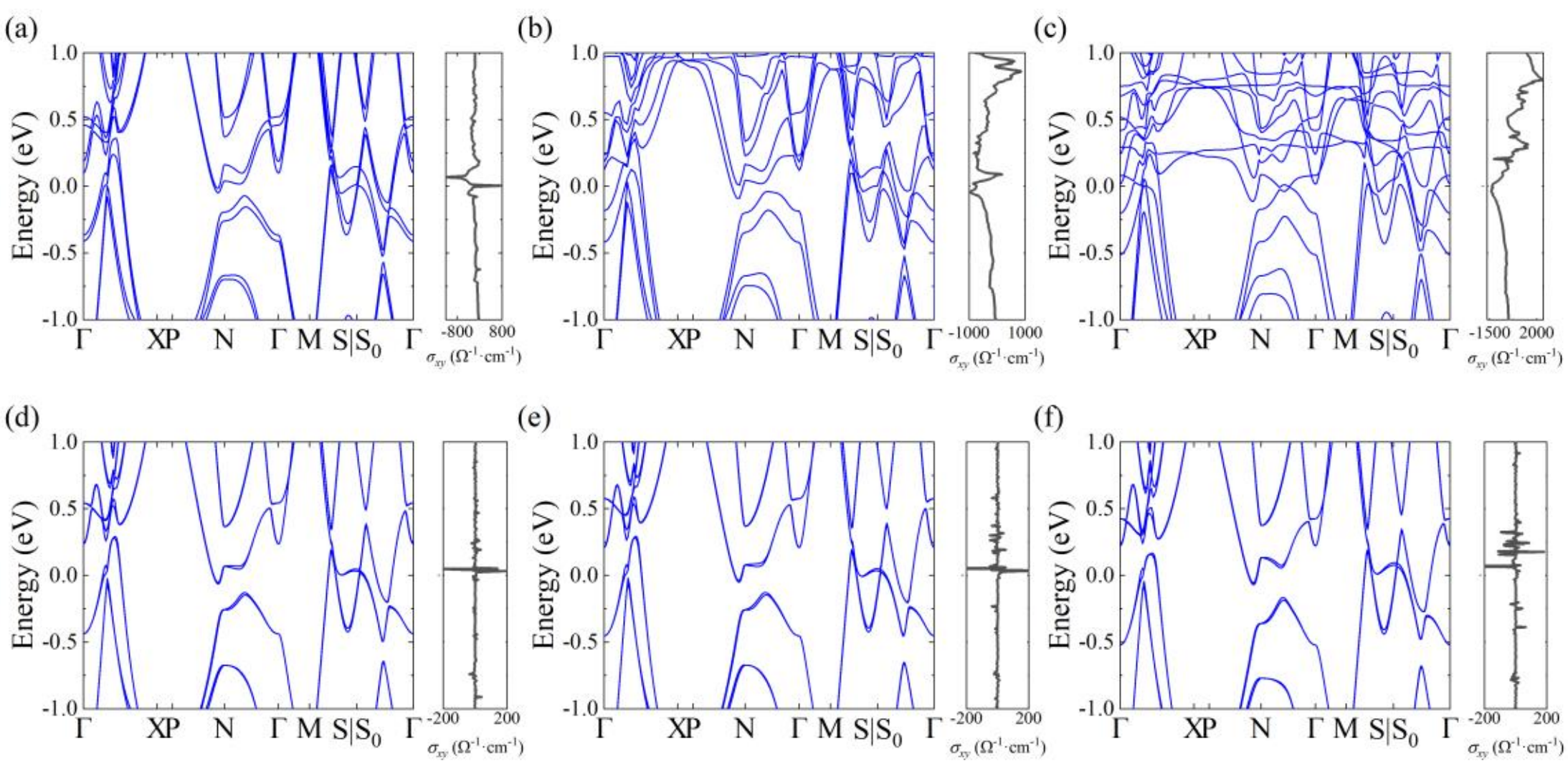

**Fig. 4.** Calculated energy bands of (a) CeGaGe, (b) PrGaGe, and (c) NdGaGe in the FM state with considering the SOC, and (d) CeGaGe, (e) PrGaGe and (f) NdGaGe in the same states without *f* valence electron. The corresponding calculated AHC as a function of the Fermi level is also presented.

To elucidate the electronic structure of the *R*GaGe family, we performed first-principles calculations including SOC. As shown in Fig. 4, the energy bands reveal that the spin splittings in PrGaGe and NdGaGe are significantly larger than in CeGaGe. Near the Fermi level, the bands in CeGaGe and PrGaGe are dominated by *d*-orbital contributions, whereas in NdGaGe, *f* orbitals also play a crucial role, suggesting a progressive increase in *f*-orbital involvement, i.e., an evolution from a dominant *d*-orbital character to a stronger *f*-electron contribution near the Fermi level. As the *R* element progresses from Ce to Pr to Nd, the *f*-electrons become increasingly dominant in the band structure, as seen by comparing the upper and lower panels of Fig. 4. The electronic structure near the Fermi level shows a stronger rare-earth dependence in *R*GaGe, with NdGaGe displaying pronounced *f*-orbital character not observed in NdAlGe, suggesting that the Ga-for-Al substitution may have altered the hybridization between rare-earth *f* states and conduction bands. Weyl points are found near the S point in all three compounds (Fig. 4), contributing substantially to the AHC. In the non-magnetic state, the breaking of inversion symmetry gives rise to multiple pairs of Weyl points [54]. In the FM state, time-reversal symmetry breaking primarily shifts the momentum-space locations of these pre-existing Weyl nodes [54, 55]; consequently, the AHE remains experimentally observable even above $T_C$. The theoretically calculated AHCs are 467, 916, and 1297 $\Omega^{-1}\cdot cm^{-1}$ for $R$ = Ce, Pr, and Nd, respectively. For comparison, our calculations for the non-magnetic state yield intrinsic AHC values of 197, 250, and 183 $\Omega^{-1}\cdot cm^{-1}$ for CeGaGe, PrGaGe, and NdGaGe, respectively. Taking PrGaGe as an example, the calculated ferromagnetic AHC (916 $\Omega^{-1}\cdot cm^{-1}$) is about three times larger than its non-magnetic value, demonstrating that magnetism significantly enhances the Berry curvature and the topological transport response. The first two values agree well with the experimental data, while the discrepancy for NdGaGe likely stems from the complex *f*-orbital physics near the Fermi level.

The band structures in Fig. 4 indicate that as magnetism increases, the band splitting induced by magnetism along the N-Γ path near the Fermi level becomes more prominent. Significantly, in NdGaGe, the conduction and valence bands gradually draw closer to each other and ultimately overlap along the N-Γ path. This phenomenon can be compared with that in $EuB_6$[56], where magnetism prompts band overlap, resulting in the formation of Weyl points. These findings indicate that the magnetism of different $R$ ions plays a crucial role in tuning the topological electronic structure.

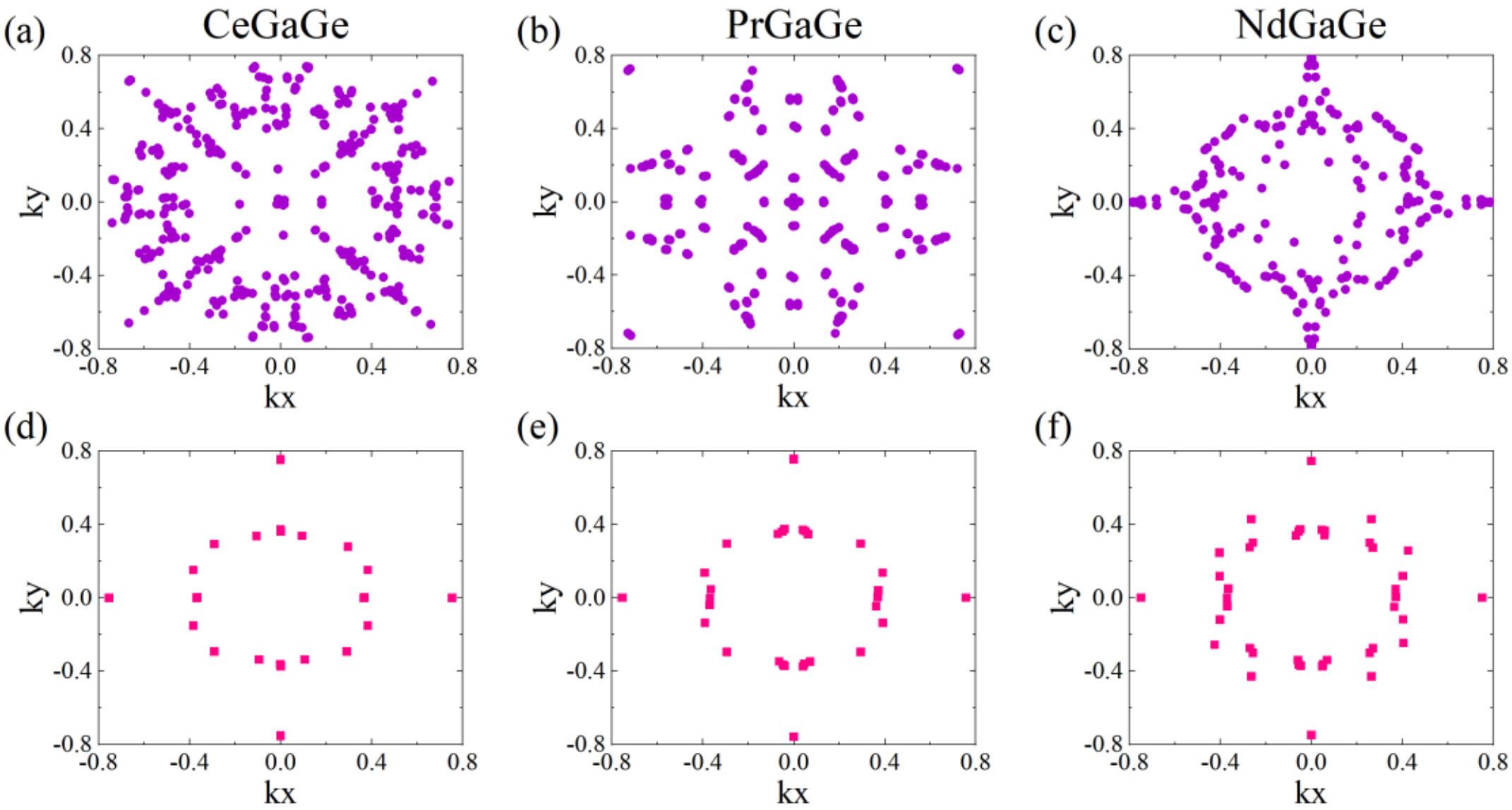

**Fig. 5**. Positions of Weyl points in the (a–c) FM and (d–f) non-magnetic phases of the *R*GaGe family.

Fig. 5 shows the positions of Weyl points in both the FM and non-magnetic phases of the RGaGe family. Most of these Weyl points do not lie along high-symmetry paths in the Brillouin zone. Spin splitting further splits the Weyl points already induced by inversion symmetry breaking. In addition, the overlap between different bands caused by spin splitting gives rise to additional Weyl points. Notably, the overlap between *f*-electron and *d*-electron bands also contributes to the presence of extra Weyl fermions in NdGaGe.

In conclusion, this comprehensive study establishes the *R*GaGe (*R* = Ce, Pr, Nd) series as a tunable platform where strong magnetism, electronic correlations, and band topology intimately interact. Using combined experimental and theoretical approaches, we demonstrate that inversion symmetry breaking stabilizes a robust Weyl semimetal state, yielding substantially enhanced intrinsic anomalous Hall conductivity compared to *R*AlGe analogs. The materials exhibit strong uniaxial magnetic anisotropy, producing distinct ground states—FM along *c*-axis and *ab*-plane AFM correlations. First-principles calculations reveal a striking rare-earth dependence of the electronic structure near the Fermi level, marked by a progressive increase in *f*-orbital contributions, evolving from dominant *d*-orbital character to substantial *f*-orbital involvement in NdGaGe. Notably, the anomalous Hall effect persists well above magnetic ordering temperatures, confirming that the topological band structure provides a robust mechanism that coexists with and is modulated by magnetism. These findings position the *R*GaGe family as a highly tunable system for investigating the interplay between localized *f*-electron physics and topological band properties, and this evolution influences the Berry curvature and anomalous Hall response while offering opportunities for realizing correlated topological phases.

*Acknowledgements.* The authors acknowledge the National Key R&D Program of China (Grant No. 2024YFA1408400 and 2023YFA140610). Y.F.G. acknowledges the open research fund of Beijing National Laboratory for Condensed Matter Physics (2023BNLCMPKF002). S.H.Z. was supported by the National Natural Science Foundation of China (12304217), the National Key Research and Development Program of China (No. 2024YFA1410300), the Natural Science Foundation of Hunan Province (No. 2025JJ60002) and the Fundamental Research Funds for the Central Universities from China (No. 531119200247). J.Y. thanks the support by the National Natural Science Foundation of China (Grants No. 12504196). The authors also thank the Analytical Instrumentation Center (#SPST-AIC10112914) and the Double First-Class Initiative Fund of ShanghaiTech University.

**References**

[1] Hsieh D, Qian D, Wray L, Xia Y, Hor Y S, Cava R J and Hasan M Z 2008 *Nature* **452** 970–4
[2] Zhang H, Liu C-X, Qi X-L, Dai X, Fang Z and Zhang S-C 2009 *Nat. Phys.* **5** 438–42
[3] Nagaosa N, Sinova J, Onoda S, MacDonald A H and Ong N P 2010 *Rev. Mod. Phys.* **82** 1539–92
[4] Yu R, Zhang W, Zhang H-J, Zhang S-C, Dai X and Fang Z 2010 *Science* **329** 61–4
[5] Wang Z, Weng H, Wu Q, Dai X and Fang Z 2013 *Phys. Rev. B* **88** 125427
[6] Xiong J, Kushwaha S K, Liang T, Krizan J W, Hirschberger M, Wang W, Cava R J and Ong N P 2015 *Science* **350** 413–6
[7] Yan B and Felser C 2017 *Annu. Rev. Condens. Matter Phys.* **8** 337–54
[8] Armitage N P, Mele E J and Vishwanath A 2018 *Rev. Mod. Phys.* **90** 015001
[9] Xu Q, Liu E, Shi W, Muechler L, Gayles J, Felser C and Sun Y 2018 *Phys. Rev. B* **97** 235416
[10] Liu D F, Liang A J, Liu E K, Xu Q N, Li Y W, Chen C, Pei D, Shi W J, Mo S K, Dudin P, Kim T, Cacho C, Li G, Sun Y, Yang L X, Liu Z K, Parkin S S P, Felser C and Chen Y L 2019 *Science* **365** 1282–5
[11] Tan W, Liu J, Li H, Guan D and Jia J-F 2022 *Quantum Front.* **1** 19
[12] Li J, Li Y, Du S, Wang Z, Gu B-L, Zhang S-C, He K, Duan W and Xu Y 2019 *Sci. Adv.* **5** eaaw5685
[13] Nakatsuji S, Kiyohara N and Higo T 2015 *Nature* **527** 212–5
[14] Bradlyn B, Cano J, Wang Z, Vergniory M G, Felser C, Cava R J and Bernevig B A 2016 *Science* **353** aaf5037

[15] Fang C, Weng H, Dai X and Fang Z 2016 *Chin. Phys. B* **25** 117106
[16] Hu J, Tang Z, Liu J, Liu X, Zhu Y, Graf D, Myhro K, Tran S, Lau C N, Wei J and Mao Z 2016 *Phys. Rev. Lett.* **117** 016602
[17] Liang T, Lin J, Gibson Q, Kushwaha S, Liu M, Wang W, Xiong H, Sobota J A, Hashimoto M, Kirchmann P S, Shen Z-X, Cava R J and Ong N P 2018 *Nat. Phys.* **14** 451–5
[18] Chi Z, Chen X, An C, Yang L, Zhao J, Feng Z, Zhou Y, Zhou Y, Gu C, Zhang B, Yuan Y, Kenney-Benson C, Yang W, Wu G, Wan X, Shi Y, Yang X and Yang Z 2018 *npj Quant Mater.* **3** 28
[19] Xia W, Bai B, Chen X, Yang Y, Zhang Y, Yuan J, Li Q, Yang K, Liu X, Shi Y, Ma H, Yang H, He M, Li L, Xi C, Pi L, Lv X, Wang X, Liu X, Li S, Zhou X, Liu J, Chen Y, Shen J, Shen D, Zhong Z, Wang W and Guo Y 2024 *Phys. Rev. Lett.* **133** 216602
[20] Yang H, Huang J, Tian S, Xia K, Wang Z, Zhang Y, Ma J, Guo H, Zhang X, Dai J, Luo Y, Wang S, Lei H and Li Y 2025 *Chin. Phys. Lett.* **42** 080706
[21] Zhong S, Orenstein J and Moore J E 2015 *Phys. Rev. Lett.* **115** 117403
[22] Sodemann I and Fu L 2015 *Phys. Rev. Lett.* **115** 216806
[23] Du Z Z, Wang C M, Sun H-P, Lu H-Z and Xie X C 2021 *Nat. Commun.* **12** 5038
[24] Hou W-T. and Zang J 2024 *Chin. Phys. Lett.* **41** 117502
[25] Xie X, Leng P, Ding Z, Yang J, Yan J, Zhou J, Li Z, Ai L, Cao X, Jia Z, Zhang Y, Zhao M, Zhu W, Gao Y, Dong S and Xiu F 2024 *Nat. Commun.* **15** 5651
[26] Wang H, and Chang K 2026 *Chin. Phys. Lett.* **43** 020703
[27] Son D T and Spivak B Z 2013 *Phys. Rev. B* **88** 104412
[28] Rong J-N, Chen L and Chang K 2021 *Chin. Phys. Lett.* **38** 084501
[29] Zhang A, Deng K, Sheng J, Liu P, Kumar S, Shimada K, Jiang Z, Liu Z, Shen D, Li J, Ren J, Wang L, Zhou L, Ishikawa Y, Ohhara T, Zhang Q, McIntyre G, Yu D, Liu E, Wu L, Chen C and Liu Q 2023 *Chin. Phys. Lett.* **40** 126101
[30] Wang J-F, Dong Q-X, Guo Z-P, Lv M, Huang Y-F, Xiang J-S, Ren Z-A, Wang Z-J, Sun P-J, Li G and Chen G-F 2022 *Phys. Rev. B* **105** 144435
[31] Lyu M, Xiang J, Mi Z, Zhao H, Wang Z, Liu E, Chen G, Ren Z, Li G and Sun P 2020 *Phys. Rev. B* **102** 085143
[32] Meng B, Wu H, Qiu Y, Wang C, Liu Y, Xia Z, Yuan S, Chang H and Tian Z 2019 *APL Mater.* **7** 051110
[33] Yang H-Y, Singh B, Gaudet J, Lu B, Huang C-Y, Chiu W-C, Huang S-M, Wang B, Bahrami F, Xu B, Franklin J, Sochnikov I, Graf D E, Xu G, Zhao Y, Hoffman C M, Lin H, Torchinsky D H, Broholm C L, Bansil A and Tafti F 2021 *Phys. Rev. B* **103** 115143
[34] Hodovanets H, Eckberg C J, Zavalij P Y, Kim H, Lin W-C, Zic M, Campbell D J, Higgins J S and Paglione J 2018 *Phys. Rev. B* **98** 245132
[35] Destraz D, Das L, Tsirkin S S, Xu Y, Neupert T, Chang J, Schilling A, Grushin A G, Kohlbrecher J, Keller L, Puphal P, Pomjakushina E and White J S 2020 *npj Quantum Mater.* **5** 5
[36] Puphal P, Pomjakushin V, Kanazawa N, Ukleev V, Gawryluk D J, Ma J, Naamneh M, Plumb N C, Keller L, Cubitt R, Pomjakushina E and White J S 2020 *Phys. Rev. Lett.* **124** 017202
[37] Wu L, Chi S, Zuo H, Xu G, Zhao L, Luo Y and Zhu Z 2023 *npj Quantum Mater.* **8** 4
[38] Su H, Shi X, Yuan J, Wan Y, Cheng E, Xi C, Pi L, Wang X, Zou Z, Yu N, Zhao W, Li S and Guo Y 2021 *Phys. Rev. B* **103** 165128
[39] Xu S-Y, Alidoust N, Chang G, Lu H, Singh B, Belopolski I, Sanchez D S, Zhang X, Bian G,

Zheng H, Husanu M-A, Bian Y, Huang S-M, Hsu C-H, Chang T-R, Jeng H-T, Bansil A, Neupert T, Strocov V N, Lin H, Jia S and Hasan M Z 2017 *Sci. Adv.* **3** e1603266

[40] He X, Li Y, Zeng H, Zhu Z, Tan S, Zhang Y, Cao C and Luo Y 2023 *Sci. China Phys. Mech. Astron.* **66** 237011

[41] Gaudet J, Yang H-Y, Baidya S, Lu B, Xu G, Zhao Y, Rodriguez-Rivera J A, Hoffmann C M, Graf D E, Torchinsky D H, Nikolić P, Vanderbilt D, Tafti F and Broholm C L 2021 *Nat. Mater.* **20** 1650–6

[42] Piva M M, Souza J C, Lombardi G A, Pakuszewski K R, Adriano C, Pagliuso P G and Nicklas M 2023 *Phys. Rev. Mater.* **7** 074204

[43] Perdew J P, Burke K and Ernzerhof M 1996 *Phys. Rev. Lett.* **77** 3865–8

[44] Blöchl P E 1994 *Phys. Rev. B* **50** 17953–79

[45] Kresse G 1995 *J. Non-Cryst. Solids* **192–193** 222–9

[46] Mostofi A A, Yates J R, Lee Y-S, Souza I, Vanderbilt D and Marzari N 2008 *Comput. Phys. Commun.* **178** 685–99

[47] Wu Q, Zhang S, Song H-F, Troyer M and Soluyanov A A 2018 *Comput. Phys. Commun.* **224** 405–16

[48] Ram D, Malick S, Hossain Z and Kaczorowski D 2023 *Phys. Rev. B* **108** 024428

[49] Yuan J, Shi X, Su H, Zhang X, Wang X, Yu N, Zou Z, Zhao W, Liu J and Guo Y 2022 *Phys. Rev. B* **106** 054411

[50] Tian Y, Ye L and Jin X 2009 *Phys. Rev. Lett.* **103** 087206

[51] Hou D, Su G, Tian Y, Jin X, Yang S A and Niu Q 2015 *Phys. Rev. Lett.* **114** 217203

[52] Wang Q, Xu Y, Lou R, Liu Z, Li M, Huang Y, Shen D, Weng H, Wang S and Lei H 2018 *Nat. Commun.* **9** 3681

[53] Sanchez D S, Chang G, Belopolski I, Lu H, Yin J-X, Alidoust N, Xu X, Cochran T A, Zhang X, Bian Y, Zhang S S, Liu Y-Y, Ma J, Bian G, Lin H, Xu S-Y, Jia S and Hasan M Z 2020 *Nat. Commun.* **11** 3356

[54] Li H 2025 *Phys. Rev. B* **112** 115134

[55] Chang G, Singh B, Xu S-Y, Bian G, Huang S-M, Hsu C-H, Belopolski I, Alidoust N, Sanchez D S, Zheng H, Lu H, Zhang X, Bian Y, Chang T-R, Jeng H-T, Bansil A, Hsu H, Jia S, Neupert T, Lin H and Hasan M Z 2018 *Phys. Rev. B* **97** 041104

[56] Gao S, Xu S, Li H, Yi C, Nie S-M, Rao Z-C, Wang H, Hu Q-X, Chen X-Z, Fan W-H, Huang J-R, Huang Y-B, Pryds N, Shi M, Wang Z-J, Shi Y-G, Xia T-L, Qian T, Ding H 2021 *Phys. Rev. X* **11** 021016